\documentclass[10pt,journal,compsoc]{IEEEtran}

\usepackage[utf8]{inputenc}%

\usepackage{tabularx,booktabs}
\usepackage{graphicx}
\usepackage{multirow}
\usepackage{acro}
\DeclareAcronym{PET}{
	short=PET,
	long=privacy-enhancing technology,
	long-plural-form=privacy-enhancing technologies
}

\ifCLASSOPTIONcompsoc
  \usepackage[nocompress]{cite}
\else
  \usepackage{cite}
\fi

\usepackage[caption=false,font=footnotesize]{subfig}

\usepackage{todonotes}
\usepackage{amssymb}
\usepackage{amsmath}

\begin{document}

\title{Using Metrics Suites to Improve the Measurement of Privacy in Graphs}

\author{Yuchen Zhao
        and~Isabel Wagner,~\IEEEmembership{Senior Member,~IEEE}%
\IEEEcompsocitemizethanks{\IEEEcompsocthanksitem Y. Zhao is with the School of Electronics and Computer Science, University of Southampton, Southampton, SO17 1BJ, United Kingdom.
\IEEEcompsocthanksitem I. Wagner (corresponding author) is with the Cyber Security Centre, De Montfort University, Leicester,
LE1 9BH, United Kingdom.\protect\\
E-mail: yuchen.zhao@soton.ac.uk, isabel.wagner@dmu.ac.uk
}%
\thanks{Manuscript received MM DD, YYYY.}}

\IEEEtitleabstractindextext{%
\begin{abstract}
Social graphs are widely used in research (e.g., epidemiology) and business (e.g., recommender systems).
However, sharing these graphs poses privacy risks because they contain sensitive information about individuals.
Graph anonymization techniques aim to protect individual users in a graph, while graph de-anonymization aims to re-identify users.
The effectiveness of anonymization and de-anonymization algorithms is usually evaluated with privacy metrics.
However, it is unclear how strong existing privacy metrics are when they are used in graph privacy.
In this paper, we study 26 privacy metrics for graph anonymization and de-anonymization and evaluate their strength in terms of three criteria: \textit{monotonicity} indicates whether the metric indicates lower privacy for stronger adversaries; for within-scenario comparisons, \textit{evenness} indicates whether metric values are spread evenly; and for between-scenario comparisons, \textit{shared value range} indicates whether metrics use a consistent value range across scenarios.
Our extensive experiments indicate that no single metric fulfills all three criteria perfectly.
We therefore use methods from multi-criteria decision analysis to aggregate multiple metrics in a metrics suite, and we show that these metrics suites improve monotonicity compared to the best individual metric.
This important result enables more monotonic, and thus more accurate, evaluations of new graph anonymization and de-anonymization algorithms.
\end{abstract}

}

\maketitle

\IEEEdisplaynontitleabstractindextext

\IEEEpeerreviewmaketitle

\IEEEraisesectionheading{\section{Introduction}}

The usage of Internet-based communications systems such as email or social networks leaves traces that can be collected and stored in graph form.
In these graphs, nodes represent users and edges represent relationships between users.
Many graph data sets have already been published for scientific or commercial use \cite{leskovec2016snap,kunegis2013konect}.
Graph data can help us understand social networks~\cite{kumar2010structure} and improve recommendations~\cite{ma2011recommender}, but can also harm user privacy because of sensitive information revealed by relationships.

To protect privacy, graphs can be anonymized by removing node identifiers and by changing the graph structure.
As a result, the nodes in the anonymized graph cannot easily be mapped to their original identifiers if the adversary does not possess additional knowledge.
However, a common assumption is that adversaries know about an auxiliary graph with similar structure as well as the correct mapping for a small number of nodes.
Finding new methods for anonymization and de-anonymization is an active research area~\cite{ji2017graph}, and researchers usually use privacy metrics to evaluate the effectiveness of their new methods.

The most commonly used metric in graph privacy is the \textit{adversary's success rate}, which gives the percentage of correctly re-identified nodes~\cite{sharad2014automated, narayanan2009deanonymizing, ji2015secgraph}.
However, even though we show in this paper that the adversary's success rate is indeed a good metric in many scenarios, it has two important shortcomings: First, it does not reveal much detail about the privacy of individual nodes because it measures privacy on a per-graph level. Second, its common definition favors de-anonymization algorithms that primarily rely on local, instead of global, graph properties.

To find a better metric for graph privacy, we analyze privacy metrics proposed in other fields~\cite{wagner2018technical} in terms of three criteria: monotonicity, evenness, and shared value range \cite{zhao2018strength}.
Monotonicity requires that privacy metrics indicate lower privacy as the adversary's strength increases.
Evenness requires that the metric values are spread evenly over their value range, which improves within-scenario comparisons.
Shared value range requires that metrics use a common value range even when they are applied to different datasets, anonymization, or de-anonymization algorithms. This improves between-scenario comparisons.

In this paper, we make the following contributions in the area of privacy measurement:

\begin{itemize}
  \item We propose a framework for the evaluation of privacy metrics for graph privacy based on our previous methodology~\cite{wagner2017evaluating,zhao2018strength}.
  \item We conduct extensive experiments to analyze the strength of 26 privacy metrics using 11 graph datasets, 6 anonymization algorithms, and 6 de-anonymization algorithms and find that no single metric excels in all three criteria.
  \item We find that several popular metrics are not monotonic in graph privacy, including entropy and the anonymity set size, and give a detailed analysis why this is the case. This finding is in contrast to results for other domains, such as vehicular network privacy~\cite{wagner2017measuring}.
  \item We propose four concrete metrics suites to improve the measurement of graph privacy and synthesize the metrics using the Weighted Product Model from multi-criteria decision analysis. We show that our metrics suites have higher monotonicity than the best individual metric.
\end{itemize}

Our findings are important because our synthesis of metrics suites shows a convenient way how the strengths of multiple privacy metrics can be combined to improve the overall measurement of privacy.

\section{Related Work}

\subsection{Graph Anonymization and De-anonymization}

Structural graph de-anonymization algorithms and corresponding anonymization algorithms have been an active research area since the mid-2000s \cite{backstrom2007wherefore}, when it was discovered that simply removing identifiers from nodes is not sufficient to prevent node re-identification.
Since then, many anonymization and de-anonymization algorithms have been proposed \cite{ji2017graph}.

In this paper, we use existing algorithms and an existing implementation \cite{ji2015secgraph} to evaluate the strength of privacy metrics for graph anonymization and de-anonymization.
We give more detail about the algorithms we used in Sections~\ref{sec:anon} and \ref{sec:deanon}.

\subsection{Privacy Metrics for Graph Privacy}
The most commonly used privacy metrics in graph anonymization and de-anonymization are the number of re-identified nodes~\cite{narayanan2008robust} and the adversary's success rate~\cite{sharad2014automated, narayanan2009deanonymizing, ji2015secgraph}.
These metrics quantify privacy as the actual privacy breach that an adversary causes.

In the wider privacy research area, other classes of privacy metrics have been proposed \cite{wagner2018technical}, for example measuring the adversary's uncertainty, information gain, or error.
For example, information-theoretic metrics such as entropy quantify privacy as the uncertainty that the adversary faces when re-identifying nodes.
These metrics have shown good strength in other fields such as vehicular networks~\cite{zhao2018strength}, but whether they are suitable for graph privacy remains unknown.
In this paper, we examine the suitability and strength of a wide range of privacy metrics in graph privacy.

De-anonymizability quantification focuses on quantifying structural properties of graph pairs, such as the edge difference, to study the maximum number of nodes that could be de-anonymized, given only structural graph information for a graph and a partially overlapping auxiliary graph \cite{ji2016relative,pedarsani2011privacy}.
However, de-anonymizability quantification does not consider the interplay between anonymization and de-anonymization algorithms and has limited applicability in concrete practical scenarios due to its focus on theoretical limits for abstract graph models \cite{ji2017graph}.

\subsection{Criteria for Privacy Metrics}

Most privacy metrics do not meet the mathematical criteria for metrics (non-negativity, identity of indiscernibles, symmetry, and triangle inequality).
Instead, several authors have proposed other criteria that good privacy metrics should fulfill.
For example, metrics should show the adversary's chances of success \cite{alexander_engineering_2003}, show the potential for privacy violations \cite{bertino_survey_2008}, measure the amount of resources needed by the adversary \cite{syverson2009why}, and integrate measurements for different aspects of privacy \cite{shokri2011quantifying}.
However, these criteria do not allow to directly compare the strength of different privacy metrics.
To this end, we have previously proposed that privacy metrics should be monotonic, i.e. that they should show lower privacy levels for stronger adversaries \cite{wagner2017evaluating,zhao2018strength}.

Most of the work that compares different privacy metrics and analyzes in which situations privacy metrics perform well is in anonymous communication.
For example, Syverson examines entropy as a metric for anonymity and concludes that it should not be used in the context of anonymous communication~\cite{syverson2009why}, and Murdoch compares the strengths and weaknesses of several different metrics for anonymous communication~\cite{murdoch2014quantifying}.

In our previous work, we have proposed a methodology to systematically evaluate different privacy metrics based on the criterion of monotonicity, and we have applied this methodology to privacy metrics for genomic privacy~\cite{wagner2017evaluating} and vehicular networks~\cite{wagner2017measuring,zhao2018strength}.
In this paper, we adapt our methodology to examine metrics in graph privacy.

\section{Methodology}
We evaluate the strength of privacy metrics for social graphs in terms of monotonicity, evenness, and shared value range.
In this section, we explain our methodology, which we have adapted from our prior work \cite{wagner2017evaluating, zhao2018strength,zhao2018evaluating}.
Our method follows six steps, as illustrated in Figure~\ref{fig:method}:

\begin{figure*}[t]
 \centering
 \includegraphics[width=\linewidth]{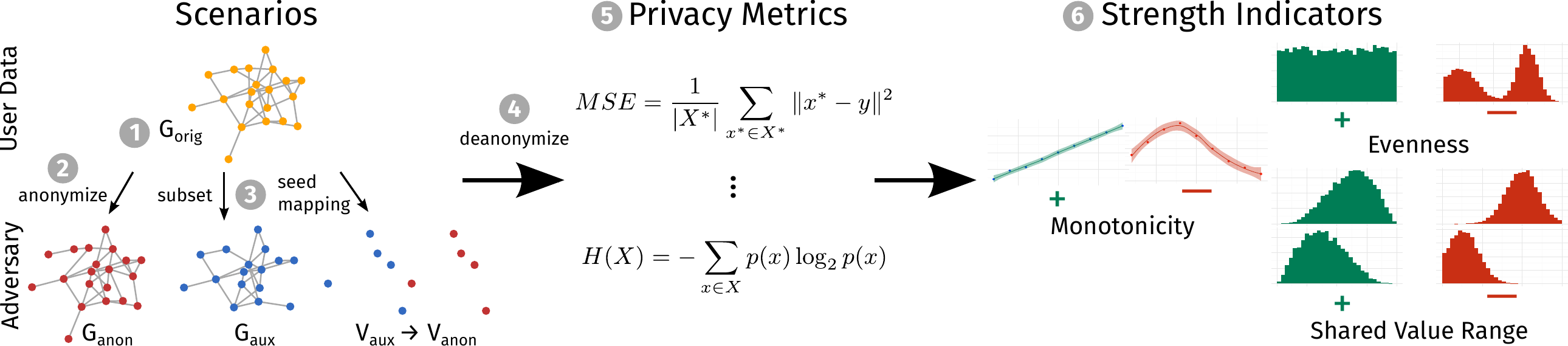}
 \caption{Our method to evaluate the strength of privacy metrics for graph privacy, adapted from prior work \cite{zhao2018strength,zhao2018evaluating}.}
 \label{fig:method}
\end{figure*} 

\begin{enumerate}
 \item Import an existing graph data set (Section \ref{sec:data})
 \item Anonymize the input graph with different anonymization algorithms (Section \ref{sec:anon})
 \item Subset the input graph to create an auxiliary graph and define seed mappings as prior information given to the adversary (Section \ref{sec:deanon})
 \item Apply different de-anonymization algorithms to the anonymized graphs, controlling the adversary's strength using the amount of prior information given to the adversary (Section \ref{sec:deanon})
 \item Compute the values of different privacy metrics based on the output of the de-anonymizer (Section \ref{sec:metrics})
 \item Analyze the strength of each privacy metric according to the criteria of monotonicity (Section \ref{sec:monotonicity}), evenness, and shared value range (Section \ref{sec:add-criteria})
\end{enumerate}

\subsection{Graph Datasets}
\label{sec:data}

We used eleven graph datasets, all available from Konect~\cite{kunegis2013konect}, to evaluate the strength of privacy metrics. Table~\ref{tab:datasets} summarizes the datasets and highlights some of the graph statistics commonly used to characterize graphs.
Our selection represents a diverse sample of graphs with respect to 19 graph statistics available in Konect.

\begin{table*}[ht]
	\caption{Characteristics of the 11 graph datasets used in our experiments, showing their diversity in several key graph statistics.}
	\label{tab:datasets}
	\centering
	\begin{tabular}{llp{1cm}p{1cm}p{1cm}p{1cm}p{1.5cm}p{1.5cm}p{1.2cm}}
		\toprule
		Dataset & Type & Nodes & Edges & Diameter & Avg. degree & Avg. shortest path & Clustering coefficient & Gini coefficient\\
		\midrule
		dblp & citation & 12591 & 49743 & 10 & 8 & 4.4 & 0.062 & 0.66 \\
		cora & citation & 23166 & 91500 & 20 & 8 & 5.9 & 0.117 & 0.52\\
		arxiv & coauthors & 18771 & 198050 & 14 & 21 & 4.2 & 0.318 & 0.61\\
		dnc & communication & 2029 & 5598 & 8 & 39 & 3.4 & 0.089 & 0.71\\
		irvine & communication & 1899 & 20296 & 8 & 63 & 3.1 & 0.057 & 0.65\\
		manufacturing & communication & 167 & 5784 & 5 & 993 & 1.9 & 0.541 & 0.44\\
		caida & computer & 26475 & 53381 & 17 & 4 & 3.9 & 0.007 & 0.63\\
		elections & online contact & 7118 & 103675 & 7 & 29 & 3.2 & 0.125 & 0.75\\
		pgp & online contact & 10680 & 24316 & 24 & 5 & 7.5 & 0.378 & 0.59\\
		google & social & 23628 & 39242 & 8 & 3 & 4.0 & 0.004 & 0.66\\
		facebook ego & social & 4039 & 88234& 8 & 2 & 3.7 & 0.519 & 0.54 \\
		\bottomrule
	\end{tabular}
\end{table*}

\subsection{Anonymizing Algorithms}
\label{sec:anon}
Graph anonymization algorithms modify the nodes or edges of a graph to prevent the individual nodes from being identified.
The simplest method is to remove the original identifiers of all nodes while leaving the graph structure intact.
This method has been shown to be susceptible to de-anonymization attacks~\cite{backstrom2007wherefore,narayanan2008robust} and we use it to represent the baseline for non-anonymized graphs.

In addition, we use five anonymization algorithms that change the graph structure: Switch~\cite{ying2008randomizing}, k-Degree Anonymity (k-DA)~\cite{liu2008towards}, Differential Privacy (DP)~\cite{sala2011sharing}, Random Walk (RW)~\cite{mittal2013preserving}, and Bounded t-Means (t-Means)~\cite{thompson2009union}.
All algorithms operate on a graph $G_{\text{orig}} = (V_{\text{orig}},E_{\text{orig}})$, where $V_{\text{orig}}$ is the set of nodes (vertices) and $E_{\text{orig}}$ is the set of edges.
We use $G_{\text{anon}} = (V_{\text{anon}},E_{\text{anon}})$ to denote the anonymized graph.

\textit{Switch} \cite{ying2008randomizing} randomly chooses two edges in $E_{\text{orig}}$ and switches them to change $G_{\text{orig}}$'s structure. 
This process is repeated $r|E_{\text{orig}}|$ times, where $r$ is the portion of edges to be switched.
The \textit{k-DA} algorithm \cite{liu2008towards} modifies the graph to achieve k-degree anonymity, so that for each node there are at least $k-1$ other nodes with the same degree.
To achieve this, the algorithm adds edges to $E_{\text{orig}}$ until the node degree for all nodes meets the requirement.
The \textit{DP} algorithm \cite{sala2011sharing} transforms the graph to a dK-series and injects noise in it.
Then the perturbed dK-series is used to re-construct a differentially private graph.
The anonymization level of DP is controlled by the parameter $\epsilon$ and the resulting graph satisfies $\epsilon$-differential privacy.
The \textit{random walk} algorithm \cite{mittal2013preserving} selects a sequence of vertices starting from $v$, for each vertex $v \in V_{\text{orig}}$, as a ``walking path'' of length $t$.
At the end of each sequence, one edge is added between the starting vertex and the ending vertex to change the graph's structure.
The \textit{t-Means} algorithm~\cite{thompson2009union} first clusters all the vertices in $V_{\text{orig}}$ into $t$ clusters based on their degree distances and then matches the degrees of vertices in a cluster to the degree of their center vertex by adding or removing edges.

To apply the anonymization algorithms in our experiments, we use the implementations provided in the open-source software SecGraph~\cite{ji2015secgraph}.

\subsection{De-anonymizing Algorithms}
\label{sec:deanon}
Graph de-anonymization attempts to re-identify nodes in an anonymized graph.
De-anonymization algorithms commonly assume that the adversary has additional knowledge in the form of an auxiliary graph $G_{\text{aux}} = (V_{\text{aux}},E_{\text{aux}})$ and a small set of seed mappings between the nodes in $V_{\text{aux}}$ and the nodes in $V_{\text{anon}}$.

The auxiliary graph is a sub-graph of the original graph, i.e. $V_{\text{aux}} \subset V_{\text{orig}}$ and $E_{\text{aux}} \subset E{\text{orig}}$, and the adversary knows the identifiers of nodes in the auxiliary graph.
In our experiments, we use the overlap of the auxiliary graph with the original graph as the first way to control the adversary's strength: a larger overlap results in a stronger adversary.

The set of seed mappings contains some relationships between nodes in the anonymized graph $G_{\text{anon}}$ and the auxiliary graph $G_{\text{aux}}$ that the adversary is assumed to know.
This set of seed mappings is typically small compared to the size of the graph, and most nodes are unmapped.
We use the number of seed mappings as the second way to control the adversary's strength: a larger number of mappings results in a stronger adversary.

The adversary's goal is to use the auxiliary graph $G_{\text{aux}}$ and the seed mappings to generate more mappings between the unmapped nodes in $G_{\text{anon}}$ and known nodes in $G_{\text{aux}}$, thus breaching graph privacy.
In our experiments, we use six de-anonymization algorithms, all implemented in SecGraph~\cite{ji2015secgraph}: Narayanan/Shmatikov (NS)~\cite{narayanan2009deanonymizing}, Ji/Li/Srivatsa/Beyah (JLSB)~\cite{ji2014structural}, Korula/Lattanzi (KL)~\cite{korula2014efficient}, Yartseva/Grossglauser (YG)~\cite{yartseva2013performance}, Adaptive De-Anonymization (ADA)~\cite{ji2014structure}, and Distance Vector based de-anonymization (DV)~\cite{srivatsa2012deanonymizing}.
Three algorithms (KL, NS, and YG) rely on seed mappings as a local property of the nodes to be matched, i.e. only inferring among the nodes that are one hop away from seed nodes, while the other three algorithms (ADA, DV, and JLSB) use properties of the global graph structure.

NS is an iterative algorithm. In each step, the adversary randomly chooses an unmapped node $v_{\text{anon}}$.
If any of $v_{\text{anon}}$'s neighbors have a known seed mapping to a node $v_{\text{aux}}$, then the set of $v_{\text{aux}}$'s neighbor nodes forms the set of candidate original nodes for $v_{\text{anon}}$.
Each candidate is scored according to its node degree, and if the node with the highest score satisfies an eccentricity criterion it is selected as the most likely original node.
If the reverse mapping from $G_{\text{aux}}$ to $G_{\text{anon}}$ of the chosen original node is $v_{\text{anon}}$, the resulting mapping is added to the set of seed mappings and the process is repeated.
The inference stops when the set of seed mappings stops growing.

The KL algorithm also bootstraps its inference from seed mappings using a heuristic called similarity witness.
The number of similarity witnesses for a pair of known and anonymized nodes is the number of seed mapping relationships between their neighbors.
In each step, the pairs with the highest number of similarity witnesses are chosen and added to the set of seed mappings.

Similarly, the YG algorithm randomly selects one unused seed mapping in each iteration and increments the scores of all its neighbor mappings (i.e. mappings between the neighbors of the known nodes and the neighbors of the anonymized node in the selected seed mapping). Once a mapping's score reaches a threshold, it is added to the set of seed mappings.

DV uses a node's distance vector, i.e. the distances between the node and the seed nodes, to describe the structural similarity between anonymized nodes and known nodes.
The algorithm maps pairs of nodes that have the highest structural similarity scores.

JLSB uses five heuristics to calculate the structural similarity between anonymized nodes and known nodes: the node degree, neighborhood, top-K reference distance, landmark reference distance, and sampling closeness centrality.
These heuristics measure both the local and global properties of each node.
The algorithm summarizes these heuristics into a structural similarity score that indicates how much an anonymized node is similar to a known node, thereby finding the most likely mappings. 

Similarly, the ADA algorithm calculates structural similarity scores from three centrality measurements: closeness centrality, betweenness centrality, and degree.
In addition, ADA takes into account the relative distance similarity (similar to the distance vector in DV) and inheritance similarity (controlling the similarity loss over iterations).

\subsection{Metrics for Graph Privacy}
\label{sec:metrics}
Based on our survey of privacy metrics \cite{wagner2018technical}, we selected 26 privacy metrics to evaluate in our experiments.
These metrics include not only metrics that have already been used in graph privacy, but also metrics from other domains.
Table~\ref{tab:metrics} summarizes the metrics we used.
According to the taxonomy in \cite{wagner2018technical}, the metrics in our study fall into five categories: uncertainty, information gain/loss, error, similarity, and success.

\begin{table}[ht]
	\caption{Graph privacy metrics used in our experiments. H/L: high (H) or low (L) values indicate high privacy. Per-graph metrics give one privacy value for the entire graph, the other metrics give one privacy value for each node. The \textit{chunk} column indicates whether a metric is negatively affected by SecGraph's chunking (see Section~\ref{sec:res-chunking}).}
	\label{tab:metrics}
	\centering
	\begin{tabular}{p{0.5cm}p{3.5cm}p{0.5cm}p{0.5cm}p{0.5cm}p{0.9cm}}
		Cat. & Metric & H/L & per-graph & gnd. truth & chunk \\
		\midrule
		\multirow{10}{*}{\rotatebox[origin=c]{90}{Uncertainty}} & Anonymity set size & H & -- & -- & $\checkmark$ \\
		& Collision entropy & H & -- & -- & $\checkmark$ \\
		& Conditional entropy & H & -- & $\checkmark$ & -- \\
		& Conditional privacy & H & -- & $\checkmark$ & -- \\
		& Entropy & H & -- & -- & $\checkmark$ \\
		& Inherent privacy & H & -- & -- & $\checkmark$ \\
		& Max-entropy & H & -- & -- & $\checkmark$ \\
		& Min-entropy & H & -- & -- & $\checkmark$ \\
		& Normalized entropy & H & -- & -- & --\\
		& Quantiles on entropy & H & -- & -- & $\checkmark$ \\
		\midrule
		\multirow{7}{*}{\rotatebox[origin=c]{90}{Information gain}} & Amount leaked information & L & $\checkmark$ & $\checkmark$ & -- \\
		& Conditional privacy loss & L & $\checkmark$ & $\checkmark$ & $\checkmark$ \\
		& Information surprisal & L & -- & $\checkmark$ & -- \\
		& Loss of anonymity & L & $\checkmark$ & $\checkmark$ & $\checkmark$ \\
		& Mutual information & L & -- & $\checkmark$ & $\checkmark$ \\
		& Pearson correlation & L & -- & $\checkmark$ & -- \\
		& Relative entropy & H & -- & $\checkmark$ & -- \\
		\midrule
		\multirow{4}{*}{\rotatebox[origin=c]{90}{Error}} & Absolute error & H & -- & $\checkmark$ & -- \\
		& Incorrectness & H & -- & $\checkmark$ & -- \\
		& Mean squared error & H & -- & $\checkmark$ & -- \\
		& \% incorrectly classified & H & $\checkmark$ & $\checkmark$ & -- \\
		\midrule
		Sim. & Normalized variance & H & -- & $\checkmark$ & -- \\
		\midrule
		\multirow{4}{*}{\rotatebox[origin=c]{90}{Success}} & Adversary's success rate & L & $\checkmark$ & $\checkmark$ & -- \\
		& Adversary's overall success & L & $\checkmark$ & $\checkmark$ & -- \\
		& Hiding property & H & $\checkmark$ & -- & -- \\
		& User-specified innocence & H & $\checkmark$ & -- & -- \\
	\end{tabular}
\end{table}

\subsubsection{Uncertainty metrics}
Uncertainty metrics measure how uncertain the adversary is about her estimate, assuming that higher uncertainty corresponds to better privacy.

For example, the \textit{anonymity set size} for each node indicates how many other nodes the adversary cannot distinguish from this node.
We approximate this notion of indistinguishability by counting how many candidate nodes have been assigned a non-zero probability.

All other uncertainty metrics in our study are based on the information theoretic concept of entropy.
\textit{Entropy} measures the adversary's uncertainty based on the probabilities assigned to each candidate node and indicates the number of additional bits of information an adversary needs to successfully de-anonymize a node.

R\'{e}nyi entropy is a generalization of entropy that introduces the parameter $\alpha$, with the entropy above using $\alpha=1$.
With increasingly larger values of $\alpha$, the influence of high-probability nodes on the metric increases.
For example, \textit{min-entropy} with $\alpha=\infty$ is based only on the node for which the adversary has the highest probability, representing the worst-case privacy.
In contrast, \textit{max-entropy} with $\alpha=0$ is based only on the number of nodes. Because max-entropy does not take into account the adversary's probabilities, it represents the best-case privacy a user can hope for.
\textit{Collision entropy} with $\alpha=2$ is another variant.

\textit{Normalized entropy} uses max-entropy to normalize entropy to a common value range of [0,1] that does not depend on the number of nodes.
\textit{Quantiles on entropy} aims to mitigate the influence of low-probability outliers on entropy, and thus computes entropy only based on a percentile of the adversary's estimated probabilities.
\textit{Conditional entropy} describes the entropy of the true mapping, conditioned on the adversary's estimate.

\textit{Inherent privacy}, or scaled anonymity set size, is based on entropy and interpreted as the number of additional yes/no questions the adversary has to answer to de-anonymize a node correctly.
\textit{Conditional privacy} is similar to inherent privacy, measuring the privacy inherent in the true mapping, given the adversary's estimate.

\subsubsection{Information gain/loss metrics}
Information gain/loss metrics focus on the amount of information that the adversary gains.

The \textit{amount of information leaked} counts how many nodes the adversary re-identified correctly. 
\textit{Pearson correlation} computes the correlation between the adversary's estimate and true node mapping. High values indicate a positive correlation and thus mean low privacy.

The remaining metrics in this category are based on information theory.
\textit{Information surprisal} focuses on the probability the adversary assigns to the true node and can be interpreted as the amount of surprise felt by the adversary on learning the true mapping.

\textit{Mutual information} indicates how much information is shared between the adversary's estimate and the true mapping, with more shared information indicating lower privacy. 
\textit{Loss of anonymity} is the maximum mutual information for any node, and thus indicates the worst-case privacy.
\textit{Conditional privacy loss} is a way of normalizing mutual information and can be interpreted as the fraction of privacy lost through the adversary's estimate.

\textit{Relative entropy}, or Kullback-Leibler divergence, measures the distance between the adversary's estimate and the true mapping, thus indicating how far the adversary's estimate is from the truth.

\subsubsection{Error metrics}
Error metrics quantify the difference between the adversary's estimate and the true mapping of candidate nodes.

\textit{Incorrectness} is the expectation of the adversary's estimation error, where successful resp. unsuccessful identification of a node is encoded as 0 resp. 1, and the expectation is computed using the adversary's estimated probabilities.

The \textit{mean squared error} measures the error between the adversary's estimated probability distribution and the true outcome.
The \textit{absolute error} measures the difference between the adversary's probability for the true node and the adversary's highest probability, i.e. for node the adversary believes to be the true node.

The \textit{percentage of incorrectly classified} nodes describes the percentage of nodes that the adversary has de-anonymized incorrectly, in relation to the total number of nodes in the graph.

\subsubsection{Similarity metrics}
Similarity metrics focus on statistical properties that measure similarity between the adversary's estimate and the ground truth.

\textit{Normalized variance} computes the variance of the difference between the true mapping and the adversary's estimate, normalized by the variance of the true mapping. A higher variance is thought to correlate with higher privacy.

\subsubsection{Success metrics}
Success metrics evaluate how likely it is that the adversary succeeds.
The \textit{adversary's success rate}, or accuracy, indicates the percentage of nodes that the adversary identified correctly.
In the variant implemented by SecGraph, this metric indicates the percentage in relation to the number of nodes that the adversary has attempted to de-anonymize, \textit{not} to the total number of nodes in the graph.

The \textit{adversary's overall success rate} indicates the success rate based on the entire graph, not just on the number of de-anonymization attempts.

The \textit{hiding property} counts the number of nodes for which the adversary's largest probability is below a specified threshold. This metric indicates the number of nodes that the anonymizing algorithm has successfully protected from the adversary.
\textit{User-specified innocence} counts the number of nodes for which the adversary's estimated probability for the true outcome is below a specified threshold, indicating the number of nodes that can reasonably claim that the adversary's de-anonymization attempt was not reliable.

\subsection{Computation of Monotonicity Scores}
\label{sec:monotonicity}
To evaluate the strength of privacy metrics, we require that they are monotonic, i.e. that they indicate higher privacy levels for stronger adversaries.
For example, we expect that, with increasing adversary strength, the values of the adversary's success rate (as a lower-better metric) decrease, and that the values of entropy (as a higher-better metric) increase.

We use the algorithm originally proposed in~\cite{wagner2017evaluating} to compute monotonicity scores.
The algorithm uses two statistical tests (t-test and rank-sum test) to compare the mean metric values for each pair of adjacent adversary strengths.
If the difference between the means is statistically significant and indicates a change in the expected direction, the algorithm increases the metrics' monotonicity score by 1.
If the difference indicates a change in the wrong direction, the algorithm subtracts 1 from the monotonicity score.
If the changes in metric values change direction, e.g. increasing for one pair and decreasing for the next, the algorithm reduces the score by 2 because such a peak may indicate the same privacy levels for both strong and weak adversaries and is thus undesirable.
A metric's final monotonicity score is the average of the scores for the two statistical tests, normalized to $[0,1]$.

\subsection{Computation of Additional Criteria}
\label{sec:add-criteria}

In addition to monotonicity, two additional criteria are useful to judge the strength of privacy metrics, especially if they will be used for within-scenario comparisons and between-scenario comparisons \cite{zhao2018strength}.

Within-scenario comparisons compare the privacy levels of different nodes within the same graph. Ideally, the values of metrics for within-scenario comparisons should be spread evenly over the entire value range (\emph{evenness}).
We compute the evenness of a metric's value range based on the Cram\'{e}r-von Mises criterion, which measures the goodness of fit between the uniform distribution U(0,1) and the normalized metric values for all adversary strengths. We normalize the Cram\'{e}r-von Mises criterion by the number of metric values to offset the influence of the number of samples.

Between-scenario comparisons compare privacy levels between different graphs, and ideal metrics use the same value range regardless of the graph's characteristics (\emph{shared value range}).
We formalize this notion by computing the portion of the global value range for each metric that is used when the metric is applied to a specific graph.

\section{Experiments}
We conducted extensive experiments to evaluate monotonicity, evenness, and shared value range in a wide range of scenarios.
In this section, we give details on the implementation and availability of software, the parameter settings we used, and how we controlled the statistical significance of our results.

\subsection{Implementation}

We have implemented our framework for the evaluation of privacy metrics in Python.
Our implementation of privacy metrics follows the description in \cite{wagner2018technical} and relies on Python libraries numpy, scipy, and scikit-learn.

For anonymization and de-anonymization algorithms, we used the open-source software SecGraph~\cite{ji2015secgraph}.
However, SecGraph's implementation of de-anonymization algorithms only outputs the adversary's node mapping and success rate.
To be able to compute other privacy metrics, we modified the de-anonymization methods to additionally output the ground truth and the adversary's estimated probability distribution.
The probability distribution is based on the adversary's scores for candidate nodes.
When the adversary tries to map a node $v_{\text{anon}}$ to a candidate in a set of candidate nodes $\{v_{\text{aux}}^1, v_{\text{aux}}^2, ...\}$, the random variable $X$ describes the probabilities for each candidate in the set.
The probability for each candidate node is defined as its score divided by the sum of all scores in the set.
In this way, the adversary creates one probability distribution for each node she attempts to re-identify.

To automate our experiments and make use of computing resources available to us, we packaged our framework in a Docker container and used Boinc and boinc2docker to distribute the computation to around 60 PCs available in student labs.
Our source code is available at CodeOcean~\cite{wagner2018privacycode}.

\subsection{Parameter Settings}

We reproduce parameter settings for the anonymization and de-anonymization algorithms as much as possible from~\cite{ji2015secgraph}.
However, the parameters given in the SecGraph paper do not always match the implementation in their software, and the software parameter choices are not documented. In these cases, we chose parameter settings that seemed reasonable to us.
Table~\ref{tab:parameters} summarizes our parameter settings.
The last two rows show the sequence of parameters we used to increase the adversary's strength.
We varied seed numbers and auxiliary ratios in independent sets of experiments and used the value given as default for the variable that was kept constant.

\begin{table}
\caption{Parameter settings used in our experiments.}
\label{tab:parameters}
\centering
 \begin{tabular}{lp{5.85cm}}
 \toprule
  \textbf{Anonymizers} \\
  DP & $\epsilon=1$ \\
  k-DA & $k=5$ \\
  RW & distance $=2$ \\
  Switch & fraction $r=0.05$ \\
  t-means & max size $=30$ \\
  
  \textbf{De-anonymizers} \\
  ADA & $\theta=0$, chunk size $=100$, $\epsilon=0.5$, weights: $w_{\text{distance}}=0.6$, $w_{\text{structural}}=0.2$, $w_{\text{inheritance}}=0.2$ \\
  DV & $\theta=0$, chunk size $=100$ \\
  JLSB & $\theta=0$, chunk size $=100$, weights: $w_{\text{degree}}=0.3$, $w_{\text{neighbor}}=0.3$, $w_{\text{ref distance}}=0.4$ \\
  KL & $\theta=1$, chunk size $=100$ \\
  NS & $\theta=0.5$ \\
  YG & $\theta=2$ \\
    
  Seed numbers & 5, 10, 20, 35, 50, 100 (default 50) \\
  Auxiliary ratios & 0.6, 0.7, 0.8, 0.85, 0.9, 0.95 (default 0.85) \\
  \bottomrule
 \end{tabular}

\end{table}

Our experiments are based on random inputs, for example the selection of seed mappings and the overlap of the auxiliary graph.
To obtain statistically significant results, we replicated the experiments until the relative error for the metric values for each combination of dataset, anonymizer, and deanonymizer was below 5\%.
100 replications were sufficient in most cases, but in some cases we computed up to 50.000 replications.

\section{Results on the Monotonicity of Graph Privacy Metrics}
\label{sec:res-monotonicity}

Our experiments have yielded results for 792 individual scenarios, i.e. 792 combinations of dataset, anonymizer, de-anonymizer, and adversary strength type, with each scenario composed of results for six adversary strength levels.
In this section, we discuss our results for monotonicity and analyze the factors that influence monotonicity. Results for the additional criteria evenness and shared value range are in the next section.

\subsection{Overview of Results}

\begin{figure}
 \centering
 \subfloat[\textit{Adversary's success rate}, DV, ID removal, Irvine, varying aux ratio. Monotonicity=1.0]{\label{fig:succ1}{\includegraphics[width=.48\linewidth]{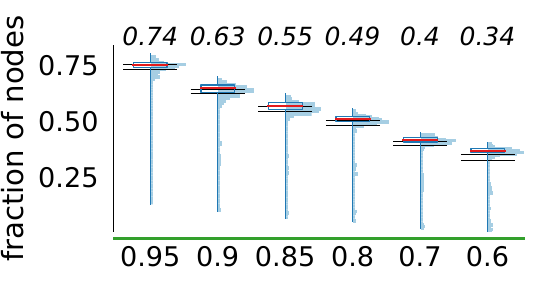} }}
 \hspace{0.05cm}
 \subfloat[\textit{Adversary's success rate}, KL, t-means, Caida, varying seed numbers. Monotonicity=1.0]{\label{fig:succ2}{\includegraphics[width=.48\linewidth]{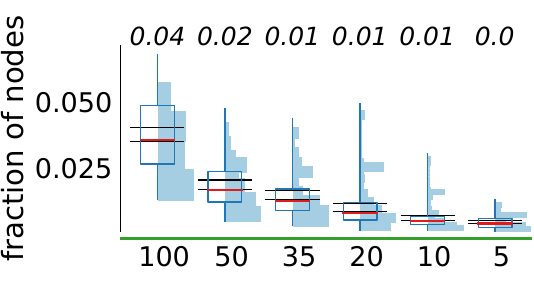} }}\\
 \subfloat[\textit{Anonymity set size}, NS, Switch, Elections, varying seed numbers. Monotonicity=0.14]{\label{fig:ass1}{\includegraphics[width=.48\linewidth]{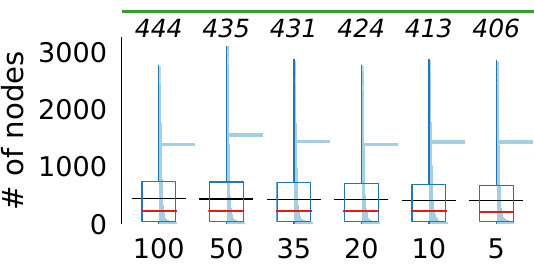} }}
 \hspace{0.05cm}
 \subfloat[\textit{Anonymity set size}, JLSB, RW, DBLP, varying aux ratio. Monotonicity=0.41]{\label{fig:ass2}{\includegraphics[width=.48\linewidth]{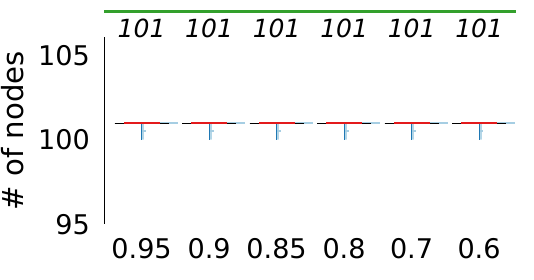} }}\\
 \subfloat[\textit{Entropy}, YG, DP, Arxiv, varying aux ratio. Monotonicity=0.46]{\label{fig:ent1}{\includegraphics[width=.48\linewidth]{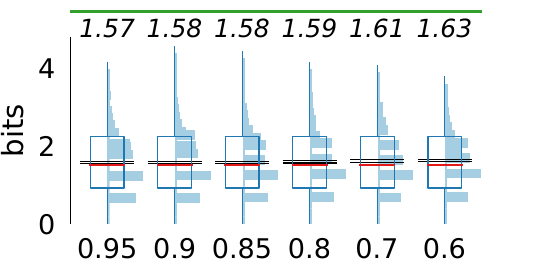} }}
 \hspace{0.05cm}
 \subfloat[\textit{Entropy}, ADA, k-DA, DNC, varying seed numbers. Monotonicity=0.11]{\label{fig:ent2}{\includegraphics[width=.48\linewidth]{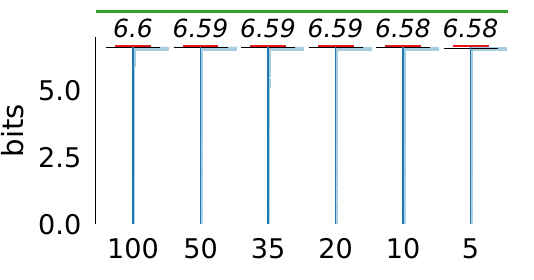} }}
 \caption{Detailed results for a selection of metrics.}
 \label{fig:violins}
\end{figure}

Figure \ref{fig:violins} shows detailed results for three metrics in six of our 792 scenarios.
Each subfigure shows of a sequence of box plots, one for each adversary strength level, with rotated histograms indicating the distribution of metric values collected from all replications.
Black horizontal lines indicate confidence intervals for the mean, italic values on top of each box indicate the mean value, and the green line at the top (resp. bottom) indicates whether higher (resp. lower) values indicate higher privacy.

According to our monotonicity criterion, we expect that boxes on the right-hand side of the plots are closer to the green line than boxes on the left, i.e. that metrics indicate higher privacy for lower adversary strengths.
The first row shows the \textit{adversary's success rate} in two scenarios (Figures~\ref{fig:succ1}--\ref{fig:succ2}). In both cases, the metric indicates a clear decrease in the success rate from left to right, with a monotonicity score of 1.0.

The plots in the second row show the \textit{anonymity set size} in two scenarios (Figures~\ref{fig:ass1}--\ref{fig:ass2}).
In Figure~\ref{fig:ass1}, the low monotonicity score results from a change in the wrong direction: instead of the \textit{anonymity set size} getting larger with decreasing adversary strength, it is actually getting smaller.

The lack of variation in Figure~\ref{fig:ass2} is due to an undocumented implementation detail in SecGraph: for all de-anonymizers that use the global graph structure (ADA, DV, JLSB), the SecGraph implementations process the graph in chunks to keep the runtime within reasonable limits (as the plot shows, we chose a chunk size of 100). This chunking does not have a large effect on the \textit{adversary's success rate} because the nodes in the anonymized and auxiliary graphs are ordered by their degrees before chunking. However, metrics that use the adversary's probabilities are skewed by this artificial limit.

The third row (Figures~\ref{fig:ent1}--\ref{fig:ent2}) shows entropy in two scenarios.
\textit{Entropy} measures the adversary's uncertainty, and we expect that \textit{entropy} increases with higher anonymization levels.
In Figure~\ref{fig:ent1}, the general trend of the metric is in the right direction. However, the monotonicity score is only a medium 0.46 because some of the adversary strength levels have no statistically significant difference.
Figure~\ref{fig:ent2} shows how \textit{entropy} is affected by SecGraph's chunking strategy: the artificial limit of 100 candidates for each node results in almost constant values for \textit{entropy} throughout.

\begin{figure}
 \centering
 \includegraphics[width=\linewidth]{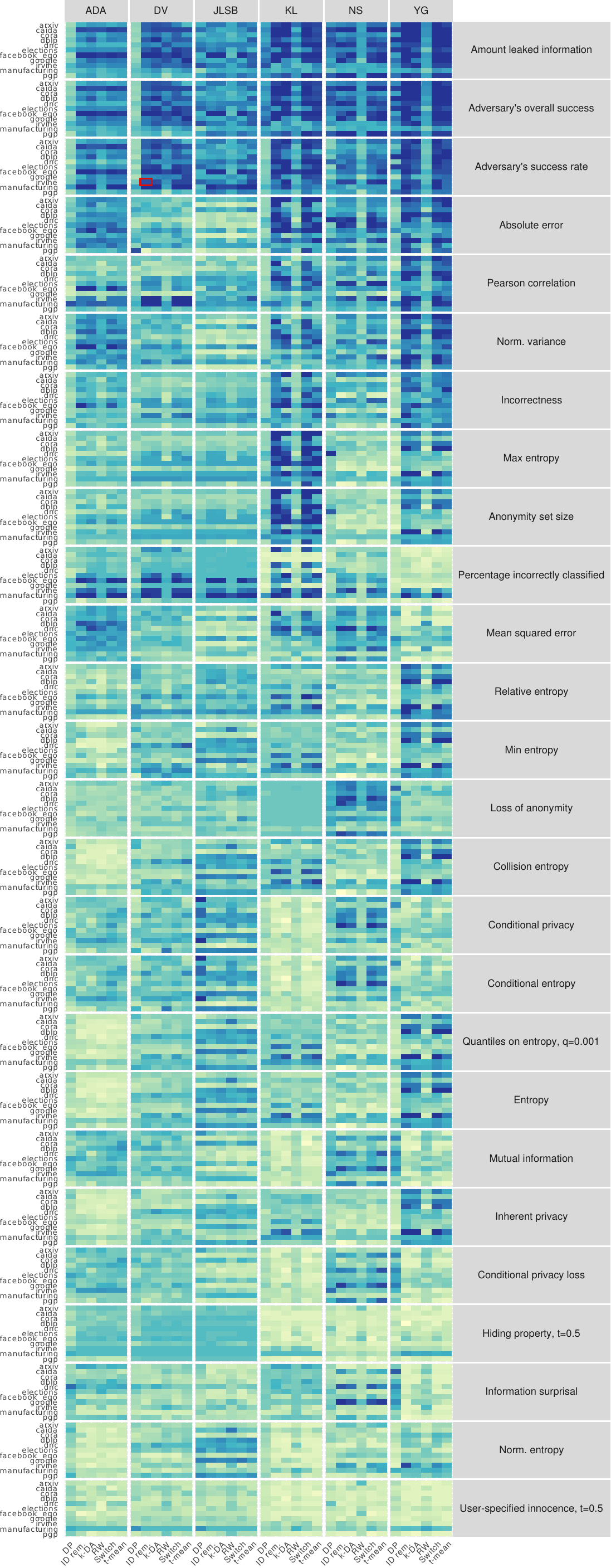}
\caption{Heat map visualizing monotonicity scores for all privacy metrics in our study, depending on the deanonymization algorithm (top), anonymization algorithm (bottom), and graph dataset (left). Each cell averages results for varying seed numbers, varying aux ratios, and all replications. Light yellow colors indicate low monotonicity (weak metric), and dark blue colors indicate high monotonicity (strong metric).}
 \label{fig:heatmap}
\end{figure}

We have applied the algorithm described in Section \ref{sec:monotonicity} to summarize the individual results from Figure \ref{fig:violins} into monotonicity scores.
Figure \ref{fig:heatmap} shows these monotonicity scores on a heat map.
Each field in the heat map represents one set of box plots from Figure \ref{fig:violins}. As an example, we have highlighted the field corresponding to Figure \ref{fig:succ1} with a red outline.
The heat map shows that only few metrics have high monotonicity, most notably the \textit{adversary's success rate} and the \textit{amount of leaked information}.
These metrics are based on the count of nodes that the adversary has successfully re-identified.

\begin{figure}[t]
 \centering
 \includegraphics[width=\linewidth]{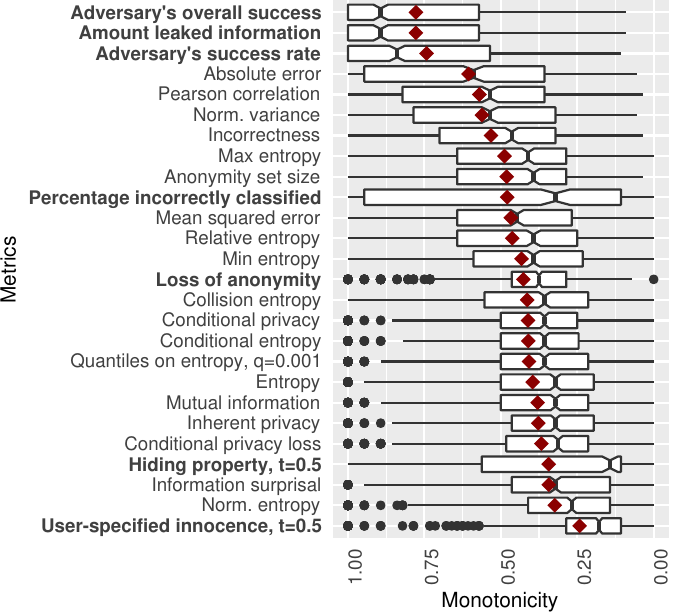}
 \caption{Box plot showing the ranking of privacy metrics according to their monotonicity scores. Each box summarizes monotonicity for all anonymization and deanonymization algorithms and all graph datasets. Per-graph metrics are marked in bold.}
 \label{fig:boxplot}
\end{figure} 

Most other metrics have medium or low monotonicity.
This includes almost all metrics that are based on the adversary's probability distribution, for example \textit{entropy} and the metrics derived from entropy.

Figure \ref{fig:boxplot} shows the monotonicity scores for all metrics in a box plot where each box combines the monotonicity scores for all anonymization algorithms, all deanonymization algorithms, and all graph datasets.
The plot is ordered according to the average score.
We can see that most metrics have monotonicity scores below $0.5$, which indicates that they are not monotonic for most anonymization levels and thus should be used with caution, if at all.
In addition, many metrics that perform well in other domains do not perform well when applied to graph privacy, for example \textit{normalized entropy} and the \textit{anonymity set size}.
We investigate the reasons for this in the following two sections.

\subsection{Analysis of Factors Influencing Monotonicity}

We have defined monotonicity as a property of privacy metrics, so we expect that the metrics should be the main determining factor for the value of monotonicity.
However, we have evaluated monotonicity in complex scenarios that include different graph datasets, anonymization algorithms, and deanonymization algorithms.
Each of these factors may influence the value of monotonicity as well.
To find out to what extent the variation in monotonicity can be explained by each of these factors, we used multivariate adaptive regression splines (MARS) \cite{friedman1991multivariate} to analyze how much each of them contributes to the final monotonicity score .
MARS is more suitable than linear regression in this case because it accounts for non-linearity and interactions between the variables and automatically selects the most important variables to include in the model.

Table~\ref{tab:mars} shows the contribution of each factor according to the MARS analysis.
The results show that metrics are indeed the most important contributors and have by far the largest influence on the monotonicity score.
In contrast, the influence from anonymizing/de-anonymizing algorithms and graph statistics is much smaller.
In many cases, the influence was so small that the variables were not important enough to be included in the model at all.

\begin{figure*}[ht]
 \centering
 \subfloat[all]{\label{fig:hist-all}{\includegraphics[width=.24\linewidth]{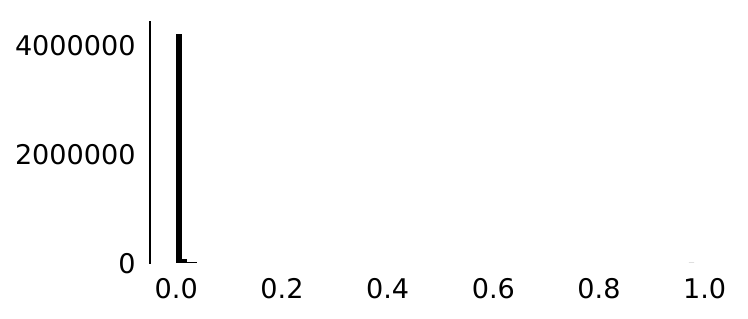} }}
 \subfloat[p=0.001]{\label{fig:hist-01}{\includegraphics[width=.24\linewidth]{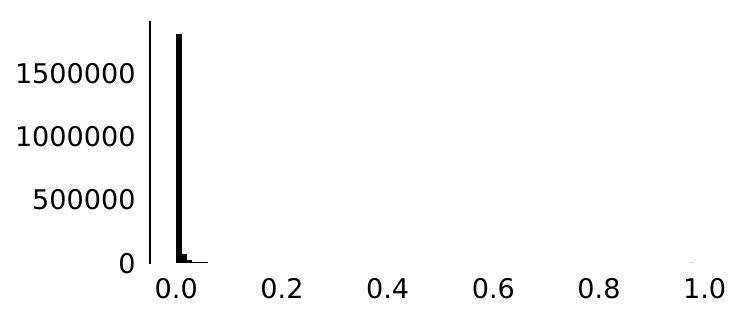} }}
 \subfloat[p=0.01]{\label{fig:hist-1}{\includegraphics[width=.24\linewidth]{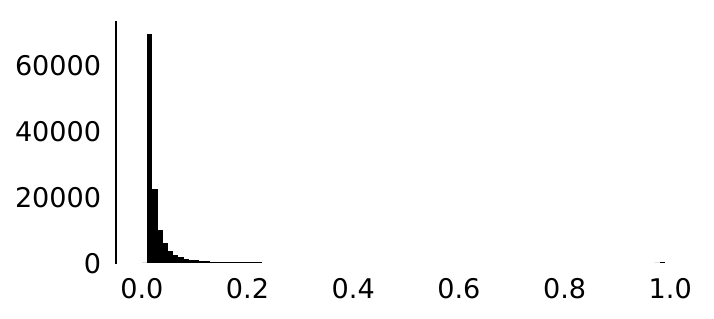} }}
\subfloat[max]{\label{fig:adv-max}{\includegraphics[width=.24\linewidth]{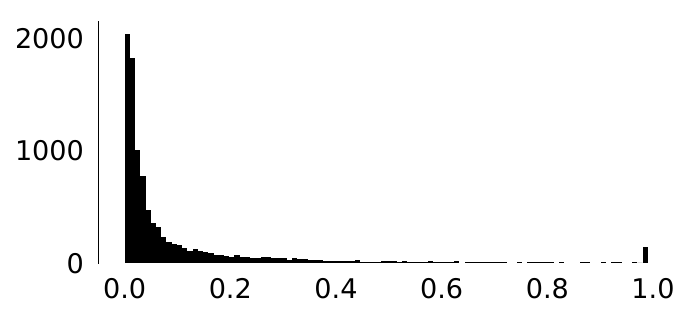} }}\\
 \caption{Distribution of the adversary's estimate for elections, NS, k-DA. Figure (a) shows all values, (b) and (c) cut off all probability values smaller than $p$, and (d) shows the probability distribution for the most likely candidate node.}
 \label{fig:histograms}
\end{figure*}

\begin{figure*}[ht]
 \centering
 \subfloat[p=0.0001, monotonicity=0.16]{\label{fig:qent-01}{\includegraphics[width=.24\linewidth]{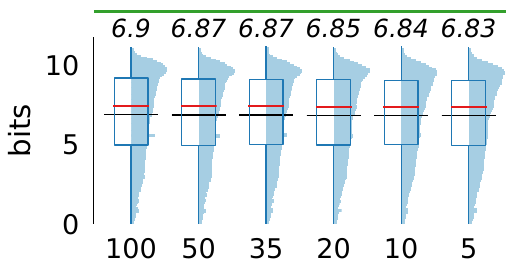} }}
 \subfloat[p=0.001, monotonicity=0.52]{\label{fig:qent-1}{\includegraphics[width=.24\linewidth]{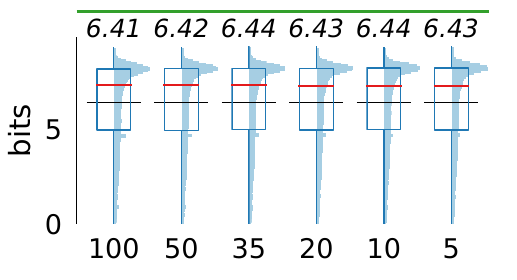} }}
 \subfloat[p=0.01, monotonicity=0.84]{\label{fig:qent-10}{\includegraphics[width=.24\linewidth]{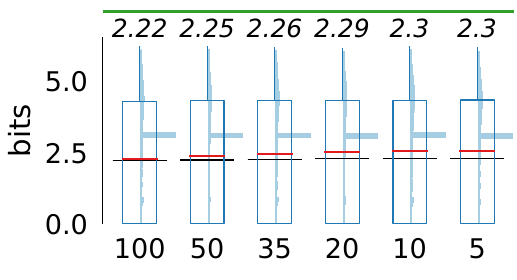} }}
 \subfloat[min-entropy, monotonicity=0.25]{\label{fig:minent-k}{\includegraphics[width=.24\linewidth]{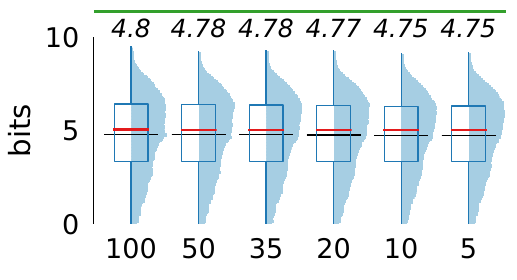} }}
 \caption{\textit{Entropy} based on the adversary's estimated probabilities without probabilities below $p$, for elections, NS, k-DA.}
 \label{fig:quantiles}
\end{figure*}

\begin{table}
\caption{MARS model showing which factors influence monotonicity, in order of importance}
\label{tab:mars}
\centering
 \begin{tabular}{llll}
\toprule
\multicolumn{4}{l}{monotonicity =}\\
\multicolumn{4}{l}{0.26} \\
  +&   0.095 &*& metric Relative entropy                                                                   \\
  +&    0.11 &*& metric Percentage incorrectly classified                                                   \\
  +&     0.2 &*& metric Pearson correlation                                                                \\
  +&    0.19 &*& metric Norm. variance                                                                     \\
  +&   0.064 &*& metric Min-entropy                                                                        \\
  +&     0.1 &*& metric Mean squared error                                                                  \\
  +&    0.12 &*& metric Max-entropy                                                                        \\
  +&   0.058 &*& metric Loss of anonymity                                                                   \\
  +&    0.16 &*& metric Incorrectness                                                                     \\
  +&    0.11 &*& metric Anonymity set size                                                                  \\
  +&    0.41 &*& metric Amount leaked information                                                           \\
  +&    0.37 &*& metric Adversary's success rate                                                            \\
  +&    0.41 &*& metric Adversary's overall success                                                         \\
  +&    0.24 &*& metric Absolute error                                                                     \\
  +&   0.089 &*& anonymizer IDrem.                                                                                \\
  +&   0.082 &*& anonymizer k-DA                                                                                  \\
  +&    0.09 &*& anonymizer Switch                                                                                \\
  +&   0.082 &*& anonymizer t-mean                                                                                \\
  +&   0.026 &*& de-anonymizer JLSB                                                                                 \\
  +&   0.037 &*& de-anonymizer YG                                                                                   \\
  -&   0.066 &*& max(0, dataset pgp -          0)  \\
  +& 5.4e-06 &*& max(0,       10680 -      nodes)  \\
  -& 5.7e-06 &*& max(0,       nodes -      10680)  \\
  -& 2.9e-06 &*& max(0,       13838 -      edges)  \\
  +& 3.6e-07 &*& max(0,       edges -      13838)  \\
  +& 8.3e-11 &*& max(0,     7.3e+08 -      claws)  \\
  +& 1.7e-11 &*& max(0,       claws -    7.3e+08)  \\
  -&  0.0083 &*& max(0,         7.7 - avg. degree) \\
  \bottomrule
  \end{tabular}
\end{table}

We fit a linear regression to the data to confirm this finding and found that all graph statistics combined account for less than 2\% of the variation in monotonicity ($R^2=0.013$), and the anonymizer, de-anonymizer, dataset and graph type combined account for less than 8\% ($R^2=0.074$).

We conclude that despite our complex evaluation setup, monotonicity is primarily influenced by the choice of metric, and not by the dataset or anonymization/de-anonymization algorithms.
Because we have evaluated a wide range of datasets, anonymizers, and de-anonymizers, we are confident that our results on the monotonicity of graph privacy metrics hold also for other datasets and algorithms.

\subsection{Analysis of Non-monotonic Metrics}

Our results show that some metrics that are popular and monotonic in other application domains are not monotonic when measuring graph privacy.
In this section we analyze why this is the case for information theoretic metrics based on \textit{entropy} and the \textit{anonymity set size}.
We also analyze the behavior of the \textit{adversary's success rate} in detail.

\subsubsection{Entropy affected by low-probability candidates}
\label{sec:res-entropy}

One reason why \textit{entropy} and metrics derived from entropy have low monotonicity scores may be that low-probability candidates have a large influence on \textit{entropy} \cite{clauss2006structuring,murdoch2014quantifying}.
To find out whether this is the case here, we plot the adversary's probability distribution using a histogram with 100 bins (Figure~\ref{fig:hist-all}).
The histogram clearly shows that most values are in the left-most bin, indicating that most probabilities are below 1\%, which confirms the presence of a large number of low-probability candidates.

As suggested in \cite{clauss2006structuring}, a possible way of dealing with these low-probability candidates is to calculate \textit{entropy} based on a quantile of the adversary's probability distribution, i.e. to remove a certain portion of low-probability candidates before calculating the entropy value.
To evaluate the effect of this, we define versions of \textit{entropy} that cut off probability values below $p$ before the calculation.
Figure~\ref{fig:quantiles} shows the detailed results and monotonicity scores for three of these modified entropies, with $p = \{0.0001, 0.001, 0.01\}$, and Figures~\ref{fig:hist-01}--\ref{fig:hist-1} show the corresponding histograms for the adversary's probability distribution.
We can see that removing probability values below 0.0001 does not improve the monotonicity score for \textit{entropy}, 0.001 shows a moderate improvement, and 0.01 shows a clear improvement.

Depending on the anonymization algorithm, removing small probability values corresponds to removing a certain percentile of the adversary's probability distribution:
removing probability values below 0.0001 corresponds to removing 0.16\% of the probability distribution, removing $p<0.001$ corresponds to removing 55\%, and removing $p<0.01$ corresponds to 97\% of the probability distribution.
Even though \textit{entropy} in the last case is monotonic, it is questionable whether it still makes sense as a privacy metric because it only evaluates the adversary's uncertainty based on the top 3\% of her probability distribution.

Another way of dealing with low-probability candidates is to use \textit{min-entropy}, a variant of \textit{entropy} that is calculated based only on the most likely candidate.
However, as Figure~\ref{fig:minent-k} shows, \textit{min-entropy} shows the same non-monotonic behavior as entropy.
To explain why, we plot the distribution of the adversary's probability for the most likely candidate for each node in Figure~\ref{fig:adv-max}.
The figure shows that for the majority of nodes, the adversary makes a mapping decision based on a probability smaller than 1\% (the left-most bin).
Just like \textit{entropy}, \textit{min-entropy} is therefore influenced by low-probability candidates.

As a result, entropy-based metrics do not seem to be suitable to evaluate graph privacy, even if only a certain percentile of the adversary's distribution is used in their calculation.

\begin{figure*}
 \centering
 \subfloat[Adversary's success rate]{\label{fig:chunk-succ}{\includegraphics[width=.24\linewidth]{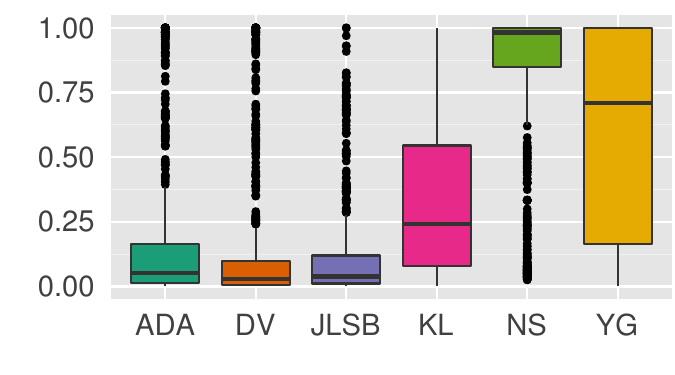} }}
 \subfloat[Anonymity set size]{\label{fig:chunk-ass}{\includegraphics[width=.24\linewidth]{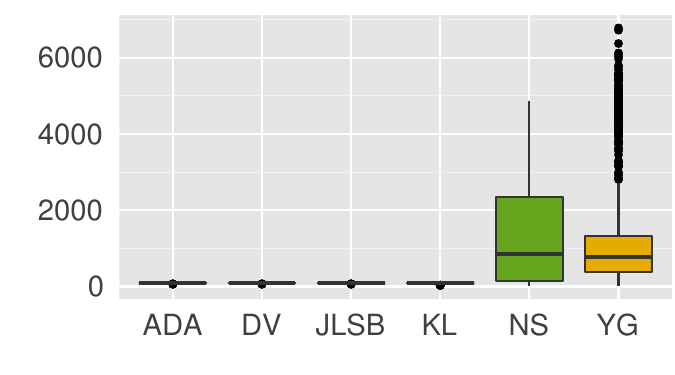} }}
 \subfloat[Mutual information]{\label{fig:chunk-mi}{\includegraphics[width=.24\linewidth]{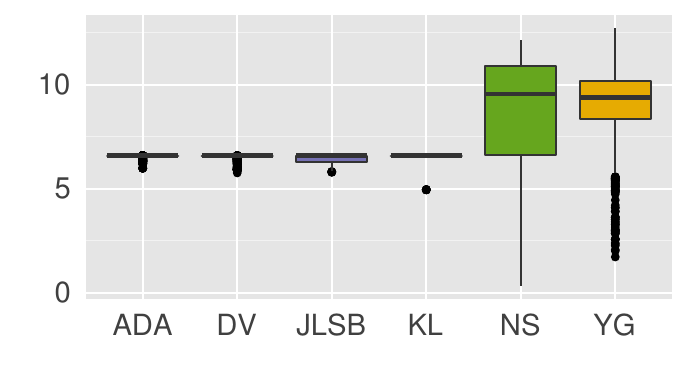} }}
 \subfloat[Entropy]{\label{fig:chunk-ent}{\includegraphics[width=.24\linewidth]{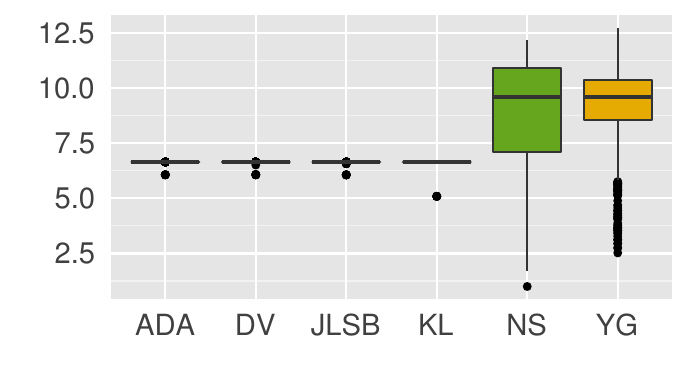} }}
 \caption{Maximum values for four metrics across all datasets and anonymizers, by de-anonymizing algorithm. The \textit{adversary's success rate} is not affected by SecGraph's chunking, whereas the other three metrics are: their maximum values have an artificial upper limit.}
 \label{fig:chunking}
\end{figure*}

\subsubsection{Metrics affected by SecGraph's chunks}
\label{sec:res-chunking}

We have shown in Figure~\ref{fig:ass2} that the anonymity set size can be influenced by SecGraph's strategy of processing graphs in chunks.
This chunking does not necessarily result in low monotonicity scores because the metric values are often indistinguishable from one adversary strength to the next, which results in a medium monotonicity.
However, the semantics and interpretation of the metric are affected: in the case of the \textit{anonymity set size}, the interpretation is no longer ``all nodes that the adversary cannot distinguish,'' but instead ``all nodes that the adversary cannot distinguish and that are in a group of 100 nodes with similar node degrees.''
This change in interpretation can make the metric unsuitable for evaluating graph privacy.

To analyze whether the \textit{anonymity set size} is the only metric affected by SecGraph's chunking, we plot the maximum values for each metric separately for each de-anonymizing algorithm. 
If a metric is affected by chunking, the four deanonymizers that use chunking (ADA, DV, JLSB, KL) should show a consistent upper limit, whereas NS and YG should not.
Figure~\ref{fig:chunking} shows these plots for four example metrics.
It is clear that the \textit{anonymity set size}, \textit{mutual information}, and \textit{entropy} are affected by chunking, but the \textit{adversary's success rate} is not.
Table~\ref{tab:metrics} (column \textit{chunk}) indicates which metrics are negatively affected by SecGraph's chunking.

\subsubsection{Adversary's Success Rate vs Overall Success Rate}

We have already mentioned the difference between the \textit{adversary's success rate} and \textit{adversary's overall success} in Section~\ref{sec:metrics}: success rate is based on the number of attempted de-anonymizations, whereas overall success rate is based on the total number of nodes in the graph.
We will show in this section why this difference can lead to misjudgment of privacy levels when comparing different de-anonymization algorithms.

We have studied two groups of de-anonymization algorithms: global algorithms (ADA, DV, JLSB) re-identify nodes based on the global graph structure, whereas local algorithms (KL, NS, YG) use only local information such as the number of neighbors.
Local algorithms generally bootstrap from the given seed mappings and iteratively add nodes to the set of seeds, stopping when no more nodes can be added.
As a result, local algorithms attempt to re-identify fewer nodes than global algorithms, and this increases their success rate (which is based on attempted nodes) compared to global algorithms.

\begin{figure}
 \centering
 \includegraphics[width=\linewidth]{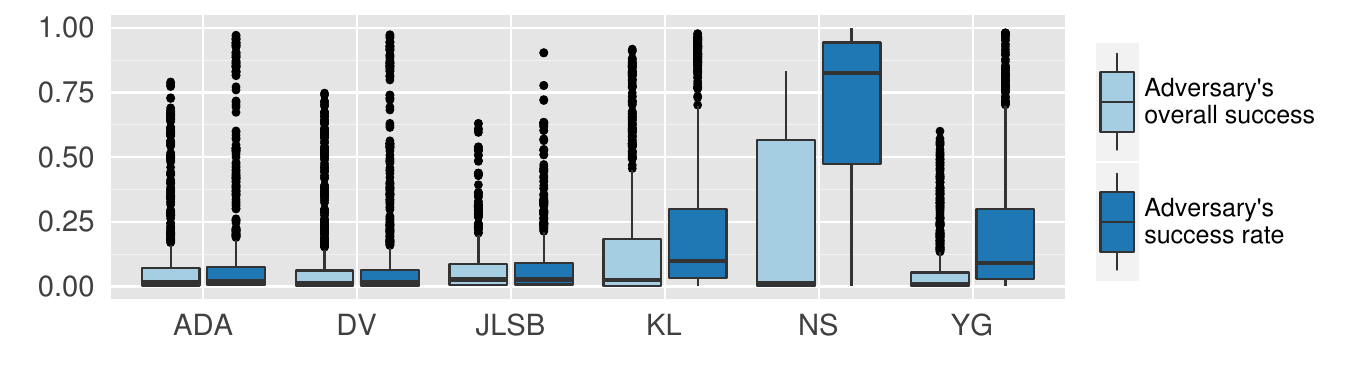}
 \caption{Comparison between the \textit{adversary's success rate} and the \textit{adversary's overall success}. De-anonymizers that use local information (KL, NS, YG) have an inflated success rate because it is normalized with the number of attempted re-identifications.}
 \label{fig:succ-overall}
\end{figure}

Figure~\ref{fig:succ-overall} compares the two metrics for each of the six de-anonymization algorithms.
We can see that the \textit{adversary's overall success} (light color) is nearly equal for all six algorithms, especially when comparing the median value indicated by the line in each box.
For the global algorithms, the \textit{adversary's success rate} is very close to the \textit{adversary's overall success}.
In contrast, the local algorithms report a much higher success rate than overall success rate.

There is no doubt that the \textit{adversary's success rate} can be a useful metric, for example to highlight the potential of a deanonymizer if the right kind of auxiliary information is available to the adversary.
However, it is unsuitable for comparing different deanonymizing algorithms because it reports inflated success rates for local algorithms.

\section{Results on Additional Criteria}

Our analysis so far has focused on the monotonicity of privacy metrics.
However, as the comparison between the \textit{adversary's success rate} and \textit{overall success rate} has shown, monotonicity is not always a sufficient criterion to select privacy metrics.
Prior work has identified evenness and shared value range as additional criteria for metric selection especially for within-scenario comparisons (evenness) and between-scenario comparisons (shared value range) \cite{zhao2018strength}.

\begin{figure}
 \centering
 \subfloat[\textit{Incorrectness}, Manufacturing, DV, DP. Evenness=0.14]{\label{fig:c1}{\includegraphics[width=.48\linewidth]{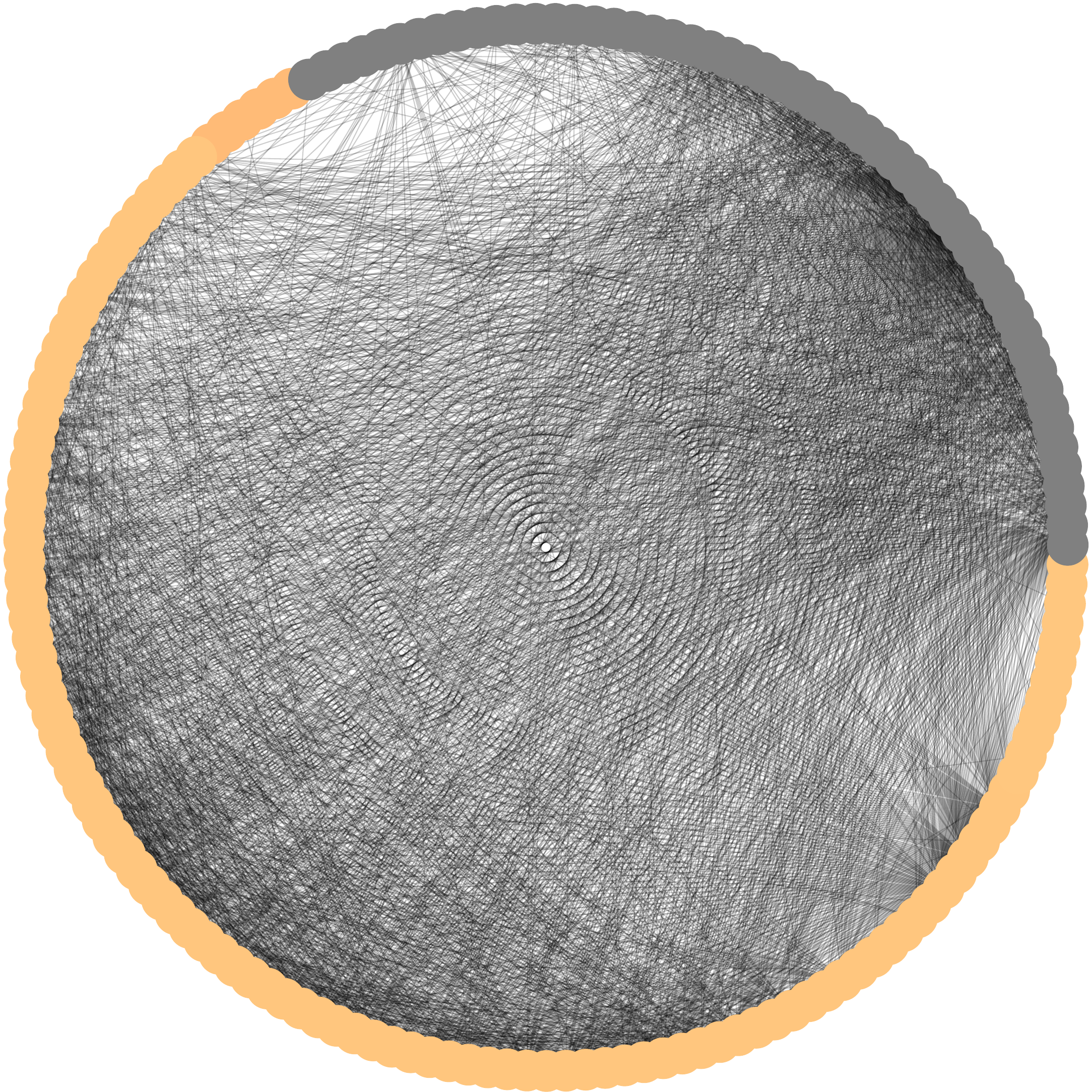} }}
 \hspace{0.05cm}
 \subfloat[\textit{Pearson correlation}, Manufacturing, DV, DP. Evenness=0.89]{\label{fig:c2}{\includegraphics[width=.48\linewidth]{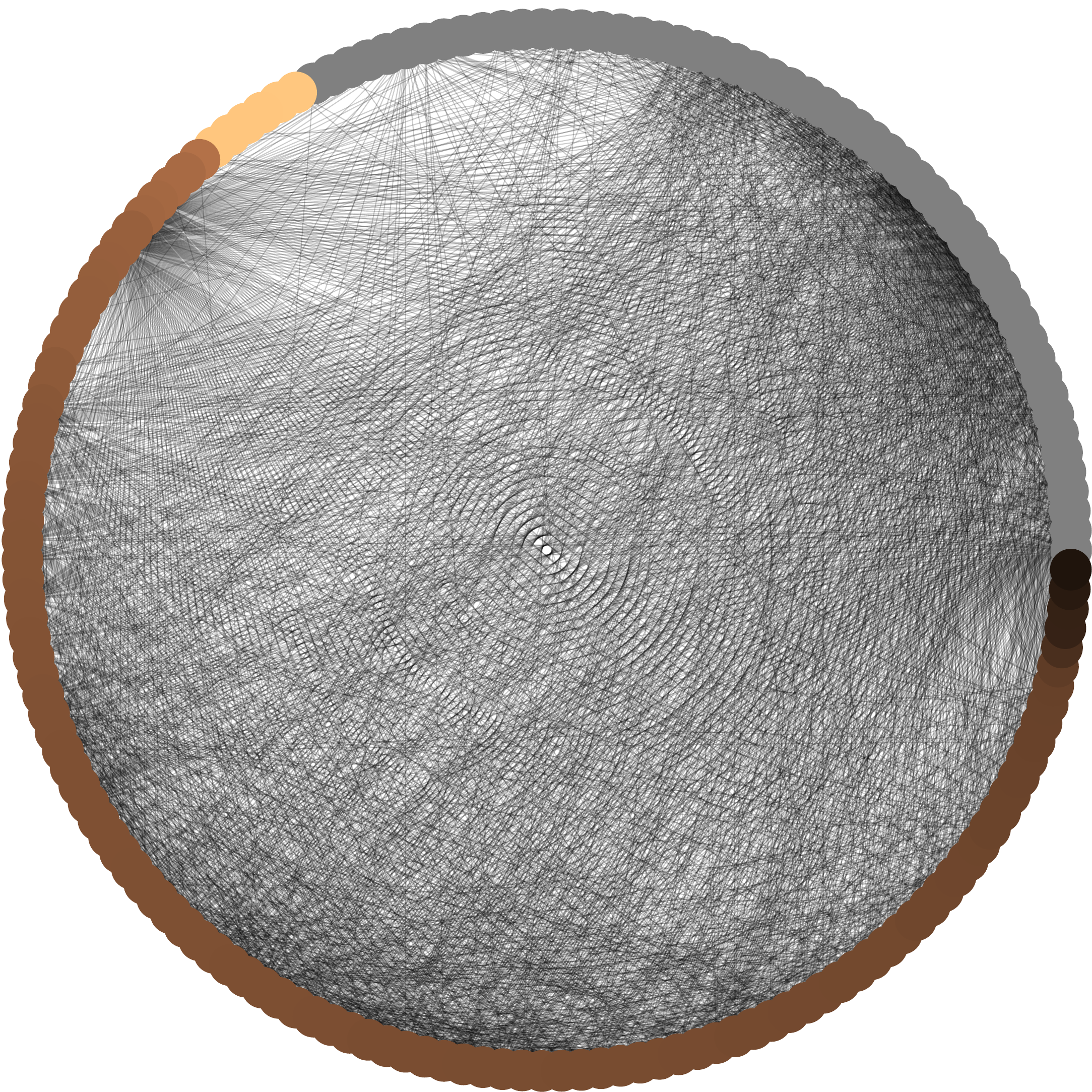} }}
 \caption{Metric values visualized on the edges of a circular graph layout. Light colors: high privacy, gray: de-anonymization not attempted.}
 \label{fig:circles}
\end{figure}

\subsection{Evenness}

To illustrate the requirement for evenness, Figure~\ref{fig:circles} compares two privacy metrics applied to the same scenario.
Each subfigure shows the graph dataset in a circular layout, with the nodes colored according to their privacy.
Light colors indicate high privacy, and nodes for which deanonymization has not been attempted are colored gray.
\textit{Incorrectness} on the left (Figure~\ref{fig:c1}) has a low evenness score (0.14), which can be seen by the absence of medium and dark colors.
In contrast, \textit{pearson correlation} on the right (Figure~\ref{fig:c2}) has a high evenness score (0.89), which is visible in the clear representation of light, medium, and dark colors.
Metrics with a high evenness score thus allow to analyze which nodes in a graph have better privacy protection than others.

\begin{figure}[t]
 \centering
 \includegraphics[width=\linewidth]{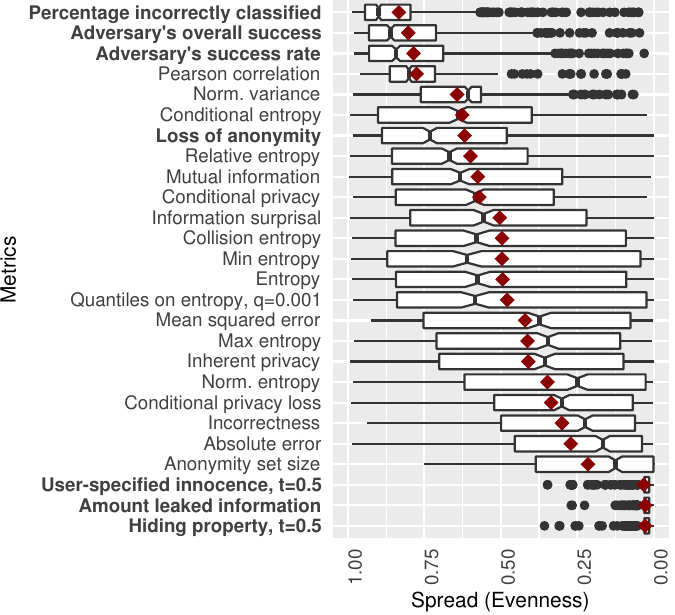}
 \caption{Box plot showing the ranking of privacy metrics according to their evenness scores.}
 \label{fig:boxplot-evenness}
\end{figure}

We have computed evenness scores according to the description in Section~\ref{sec:add-criteria} for all 792 scenarios. Figure~\ref{fig:boxplot-evenness} shows a ranking of privacy metrics according to their average evenness score, with the boxes indicating the distribution of scores across all scenarios.
Among the metrics with an average evenness score above 0.5, four metrics also have a monotonicity score above 0.5: \textit{adversary's overall success}, \textit{adversary's success rate}, \textit{pearson correlation}, and \textit{normalized variance}.

\subsection{Shared Value Range}

\begin{figure}
 \centering
 \subfloat[\textit{Pearson correlation}, Caida, varying aux ratio, YG, ID rem., range=1.0]{\label{fig:r1}{\includegraphics[width=.48\linewidth]{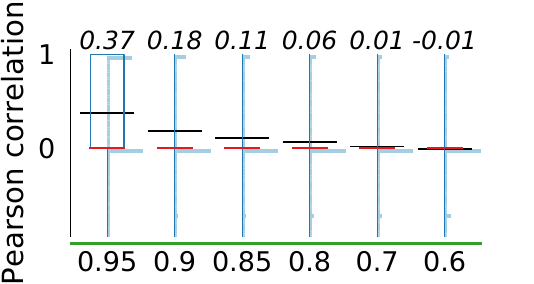} }}
 \hspace{0.05cm}
 \subfloat[\textit{Pearson correlation}, Manufacturing, varying aux ratio, JLSB, t-means, range=1.0]{\label{fig:r2}{\includegraphics[width=.48\linewidth]{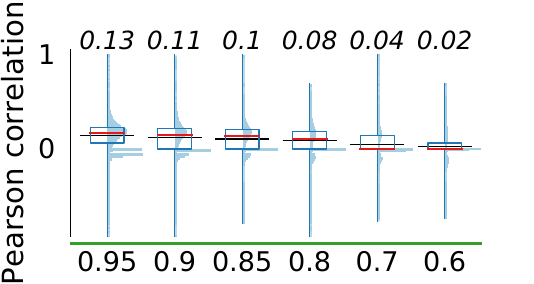} }}\\
 \subfloat[\textit{Anonymity set size}, Caida, varying aux ratio, YG, ID rem., range=0.59]{\label{fig:r5}{\includegraphics[width=.48\linewidth]{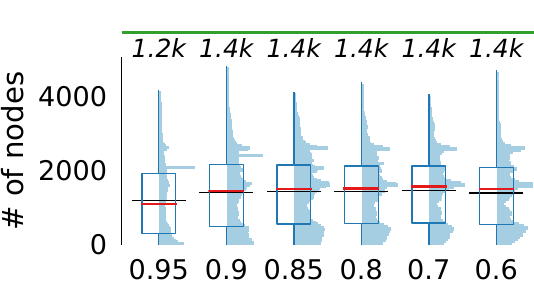} }}
 \hspace{0.05cm}
 \subfloat[\textit{Anonymity set size}, Manufacturing, varying aux ratio, JLSB, t-means, range=0.01]{\label{fig:r6}{\includegraphics[width=.48\linewidth]{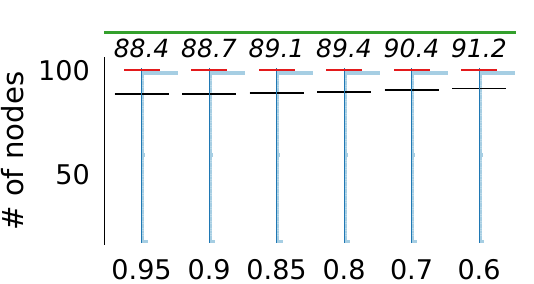} }}
 \caption{Shared value range: \textit{pearson correlation} in the top row has a consistent value range across scenarios, while the \textit{anonymity set size} in the bottom row uses a different value range (y axis) for each scenario.}
 \label{fig:violins-range}
\end{figure}

To illustrate the requirement for a shared value range, we show two metrics, \textit{pearson correlation} and \textit{anonymity set size}, in four scenarios each in Figure~\ref{fig:violins-range}.
The top row shows \textit{pearson correlation} with a consistent value range of [0,1] in every scenario.
In contrast, the bottom row shows the \textit{anonymity set size}, where each scenario uses a different portion of the global value range, depending on the size of the graph, the anonymization/de-anonymization algorithms and how they are implemented.
This indicates that the \textit{anonymity set size} is less suitable for comparisons between scenarios than \textit{pearson correlation}.

\begin{figure}[t]
 \centering
 \includegraphics[width=\linewidth]{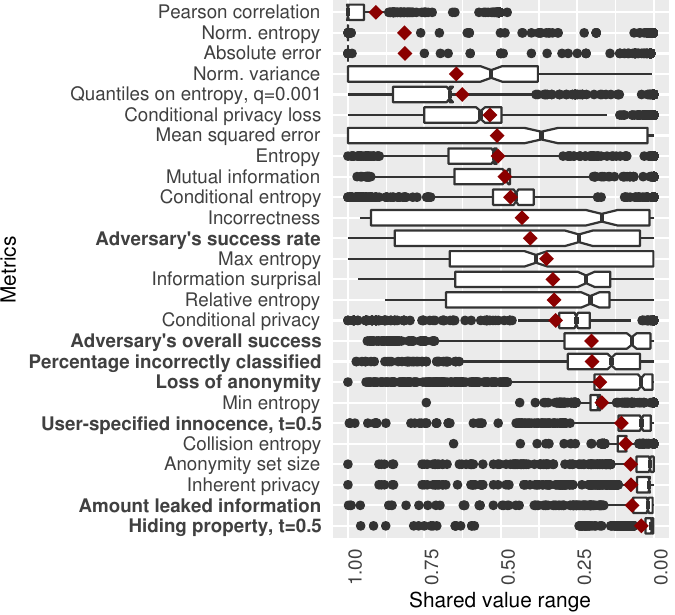}
 \caption{Box plot showing the ranking of privacy metrics according to their shared value range scores.}
 \label{fig:boxplot-range}
\end{figure}

Figure~\ref{fig:boxplot-range} shows the distribution and average values for the shared value range score across all scenarios.
Only three metrics have a shared value range score above 0.5 as well as a monotonicity score above 0.5: \textit{pearson correlation}, \textit{absolute error}, and \textit{normalized variance}.

\section{Recommendations: Metrics Suites}
Based on our experimental results and the analysis in the previous sections, we can see that no single metric is the ideal metric for all scenarios.
The metrics with the highest monotonicity are per-graph metrics, so they do not provide information about the privacy levels of individual nodes in a graph.
Most of the other metrics have very low monotonicity, and only seven metrics have an average monotonicity score of 0.5 or above.
When considering evenness and shared value range in addition to monotonicity, only two metrics have high scores in all three criteria (\textit{pearson correlation} and \textit{normalized variance}).
Finally, another important consideration is the semantics of metrics and their interpretation. In this regard, it is desirable to measure different aspects of privacy, for example in terms of the categories described in Section~\ref{sec:metrics}.

Therefore, our recommendation is to combine several privacy metrics in a metrics suite.
To allow for easy comparisons between anonymizing or de-anonymizing algorithms, and to be able to decide which provides the best or worst privacy levels, it is desirable to combine the metrics in a suite into a single number.
This is not a straightforward task because metrics have different scales of measurement and different directions (for some, higher values indicate higher privacy, for others, lower values).
A simple average is therefore unlikely to yield good results, and any normalization has to carefully consider direction and value range.

\subsection{Combining Metrics in a Metrics Suite}

The problem of combining metrics is similar to a problem in operations research: given a set of alternatives and a number of criteria for each alternative, which alternative is best?
Many methods have been proposed for multi-criteria decision analysis \cite{triantaphyllou2000multicriteria}.
The simplest method is the weighted sum model (WSM), which ranks alternatives according to a weighted sum of the criteria values.
However, this method is only applicable if the criteria are measured on the same scale and in the same units.
The weighted product model (WPM) instead computes the relative importance $Q_i$ of each alternative $i$ using a weighted product.
Applied to the case of privacy metrics, each criterion $j$ corresponds to one privacy metric, and the set of alternatives is the set of scenarios (e.g., anonymization algorithms or parameter settings) that need to be compared in terms of their privacy.
According to \cite{triantaphyllou2000multicriteria}, the relative importance of each alternative $i$ is computed as
\[
 Q_i = \prod_{j=1}^n(\overline{x}_{ij})^{w_j},
\]
where $w_j$ is the weight assigned to each metric, and $\overline{x}_{ij}$ is a normalization of the metric value $x_{ij}$ such that
\[
 \overline{x}_{ij} = \frac{x_{ij}}{\max_i x_{ij}}
\]
if higher values indicate higher privacy, and
\[
 \overline{x}_{ij} = \frac{\min_i x_{ij}}{x_{ij}}
\]
if lower values indicate higher privacy.
The advantages of WPM are that different units of measurement are effectively eliminated due to the use of multiplication instead of addition, that the built-in normalization can integrate higher-better and lower-better metrics, and that it does not suffer from rank reversals that can occur with additive methods \cite{tofallis2014add}.

\subsection{Choosing Metrics for a Metrics Suite}
As we have already discussed above, every metric in a metrics suite should have a monotonicity score of at least 0.5.
The choice of metrics can then depend on the specific demands of the application and might include metrics with high evenness, metrics with high shared value range, metrics from different categories, and metrics that are easy to interpret in the application context.

In our experiments, the seven metrics that have high monotonicity scores are the \textit{adversary's overall success}, \textit{amount leaked information}, \textit{adversary's success rate}, \textit{absolute error}, \textit{pearson correlation}, \textit{normalized variance}, and \textit{incorrectness}.
This list includes metrics with high evenness and shared value range, and cover fours different categories (information gain/loss, error, similarity, success probability). The list also includes the \textit{adversary's success rate} which is the most common metric to evaluate graph privacy.

\subsection{Evaluating the Performance of Metrics Suites}
To find out which combination of metrics results in the best metrics suite, we have evaluated all metrics suites resulting from combinations of the top-7 monotonic metrics.
A good metrics suite should create a monotonic ranking of alternatives.
Based on our experimental data, we can evaluate the monotonicity of metrics suites as follows.
Each of our 792 scenarios consists of six adversary strength levels, with increased strength defined either by an increase in the number of seed mappings or by an increase in the overlap of the auxiliary graph.
We can therefore use each scenario as one set of six alternatives, using the increasing strength levels as the ground truth for how the alternatives should be ranked.
For each set of alternatives, we used the mean metric values across all replications as $x_{ij}$ and then count how many of the six strength levels have been ranked monotonically by the metrics suite score $Q_{i}$.

We find that most metrics suites create monotonic rankings for more than 80\% of scenarios, and only metrics suites that include none of the top-3 metrics create monotonic rankings for less than 70\% of scenarios.
In comparison, the best individual metrics (\textit{adversary's overall success} and \textit{amount information leaked}) create monotonic rankings for 88.2\% of scenarios.

Table~\ref{tab:suites} summarizes the composition and weights for four of the top-scoring metrics suites and the best individual metric. 
When choosing all weights $w_j$ to be equal, the best metrics suite, consisting of \textit{pearson correlation}, \textit{adversary's overall success}, \textit{normalized variance} and \textit{amount leaked information}, ranks 88.6\% of scenarios monotonically.
Importantly, this is better than the result for the best individual metric.

When choosing unequal weights, the best metrics suite we have been able to construct consists of \textit{pearson correlation}, \textit{normalized variance}, \textit{incorrectness}, \textit{amount leaked information}, \textit{adversary's success rate}, \textit{adversary's overall success}, and \textit{absolute error} with weights $w_i = 0.1, 0.1, 0.1, 0.25, 0.1, 0.25, 0.1$. 
This metrics suite creates monotonic rankings for 89\% of scenarios.

In summary, we find that combining privacy metrics in a metrics suite can improve monotonicity above the monotonicity of individual metrics.
This is an important result that shows a concrete method towards more monotonic and thus more accurate evaluations of privacy.

\begin{table}
 \caption{Composition of four metrics suites (S2 to S5) and the percentage of monotonic rankings resulting from their aggregation with WPM, compared with the best individual metric (S1).}
 \label{tab:suites}
 \begin{tabular}{lp{4.9cm}p{1.05cm}p{1.05cm}}
  & Metrics & Weights & \% mono\\
  \midrule
  S1 & Adversary's overall success & equal & 88.2 \\
  S2 & Adversary's overall success, Norm. variance & equal & 88.6 \\
  S3 & Pearson correlation, Adversary's overall success, Norm. variance, Amount leaked information & equal & 88.6 \\
  S4 & Pearson correlation, Adversary's overall success, Norm. variance, Amount leaked information, Incorrectness & 0.1, 0.35, 0.1, 0.35, 0.1 & 88.7 \\
  S5 & Pearson correlation, Norm. variance, Incorrectness, Amount leaked information, Adversary's success rate, Adversary's overall success, Absolute error & 0.1, 0.1, 0.1, 0.25, 0.1, 0.25, 0.1 & 89.0 \\
 \end{tabular}
\end{table}

\section{Conclusion}
In this paper, we analyzed the strength of 26 privacy metrics for graph privacy in terms of their monotonicity, evenness, and shared value range.
We conducted extensive experiments on 11 public graph datasets, using 6 anonymization algorithms and 6 de-anonymization algorithms.
We found that most metrics are not monotonic when applied in graph privacy, including several metrics that are popular and monotonic in other fields.
Our detailed analysis of the strengths and weaknesses of these metrics led us to propose metrics suites, that is, combinations of privacy metrics that can combine the strengths and mitigate the weaknesses of individual metrics.
To the best of our knowledge, we were the first to apply techniques from multi-criteria decision analysis to privacy measurement and found that the resulting metrics suites can indeed increase monotonicity above the monotonicity of the best individual metric.
This result opens up a new line of research that may lead to significant improvements in privacy measurement.

\ifCLASSOPTIONcompsoc
  \section*{Acknowledgments}
\else
  \section*{Acknowledgment}
\fi
This work was supported by the UK Engineering and Physical Sciences Research Council (EPSRC) grant EP/P006752/1 and used the ARCHER UK National Supercomputing Service (http://www.archer.ac.uk).

\bibliographystyle{IEEEtranS}
\bibliography{IEEEabrv,graph-privacy-metrics}

\vfill

\end{document}